\documentclass[conference]{IEEEtran}
\IEEEoverridecommandlockouts

\usepackage{cite}
\usepackage{amsmath,amssymb,amsfonts}
\usepackage{algorithmic}
\usepackage{graphicx}
\usepackage{textcomp}
\usepackage{xcolor}
\usepackage[colorlinks,urlcolor=blue,linkcolor=blue,citecolor=blue]{hyperref}

\usepackage{color,array}
\usepackage{amsmath}
\usepackage{amssymb}
\usepackage{tabularx}
\usepackage{multirow, makecell}
\usepackage{booktabs}
\usepackage{orcidlink}
\usepackage{newtxmath}
\usepackage[utf8]{inputenc}

\usepackage{authblk}

\usepackage[ruled]{algorithm2e}

\usepackage{microtype}
\usepackage[symbol]{footmisc}
\usepackage[colorinlistoftodos,prependcaption,textsize=tiny]{todonotes}

\newcolumntype{Y}{>{\centering\arraybackslash}X}

\newcommand\mone{$\text{C}_\text{g}$}
\newcommand\mtwo{$\text{C}_\text{s}$}
\newcommand\mthree{P}
\newcommand\mfour{$\text{P}_{\text{e}}$}
\newcommand{\best}[1]{\textbf{#1}}

\newcommand{\obj}[1]{\textsl{\textsf{#1}}}

\usepackage{xifthen}
\usepackage[normalem]{ulem}
\def\clap#1{\hbox to 0pt{\hss #1\hss}}%
\def\initials#1{\protect\clap{\smash{\raisebox{1.6ex}{\tiny{\textsf{\textit{~~#1}}}}}}}%
\newcommand{\EDITbyauthorbold}[4][]{%
\strut{{\color{gray}\hspace{0pt}\initials{\color{#3}#2}\ifthenelse{\isempty{#1}}{\protect{\color{#3}\fontseries{sb}\selectfont{#4}}}{\protect\sout{#1} {\color{#3}\fontseries{sb}\selectfont{#4}}}}}}{}
\definecolor{authors}{RGB}{191, 0, 64}
\newcommand{\EDIT}[4][]{{#4}}
\newcommand{\VWedit}[2][]{\protect{}\EDIT[#1]{}{authors}{#2}}

\begin{document}

\title{MM-2FSK: Multimodal Frequency Shift Keying for Ultra-Efficient and Robust High-Resolution MIMO Radar Imaging} 

\author[1]{\href{mailto:vanessa.wirth@fau.de}{Vanessa~Wirth}\orcidlink{0000-0001-8295-3021}}
\author[3]{Johanna~Bräunig\orcidlink{0000-0003-2276-272X}}
\author[2]{Martin~Vossiek\orcidlink{0000-0002-8369-345X}}
\author[1,4]{Tim~Weyrich\orcidlink{0000-0002-4322-8844}*}
\author[1]{Marc~Stamminger\orcidlink{0000-0001-8699-3442}*}

\affil[1]{Visual Computing Erlangen, Friedrich-Alexander-Universität Erlangen-Nürnberg, Germany}
\affil[2]{Institute of Microwaves and Photonics, Friedrich-Alexander-Universität Erlangen-Nürnberg, Germany}
\affil[3]{fiveD GmbH, Germany}
\affil[4]{University College London (UCL), United Kingdom}

\maketitle

\begin{abstract}

\VWedit{
Accurate reconstruction of dynamic targets demands three-dimensional imaging at high temporal resolution.
Radar is a particularly attractive sensing modality for this purpose due to its fast capture capabilities and robustness under varying lighting and weather conditions.
Consequently, it has been successfully employed in  a variety of dynamic applications, such as automotive and robotic perception.
Nevertheless, dynamic sensing remains comparatively unexplored in operating regimes requiring near-field, high-fidelity surface reconstruction, as commonly targeted by large-aperture mmWave MIMO radars. 
Here, achieving high frame rates is challenged by the substantial computational cost of reconstruction, imposing stringent constraints on bandwidth, compute resources, and ultimately sensor capture rate.}
\VWedit{To enable fast captures at reduced computation cost, prior work explored few-tone imaging, reducing the number of transmitted frequencies at the expense of increased range-phase ambiguity;} however, this requires coarse knowledge of the target position \VWedit{and is primarily designed to reconstruct flat surface geometries}.
\VWedit{We present MM-2FSK, which extends two-tone stepped frequency imaging with an external modality-agnostic per-point depth prior for precise near-field 3D imaging of geometrically complex surfaces.
To address range-phase ambiguities, the prior is processed to mitigate inter-modality data mismatch, transformed into a radar-compatible representation, and integrated into signal processing  as a phase constraint.
In experiments with a mmWave FSCW MIMO radar and with representative priors obtained from a co-located optical sensor, we demonstrate superior depth accuracy of our method over comparable near-field few-tone radar imaging approaches in static and dynamic scenes, competitive with time- and resource-intensive measurements of many-tone backprojection. }

\end{abstract}

\begin{IEEEkeywords}
3D reconstruction, depth cameras, frequency shift keying, mimo radar, multimodal, radar imaging, sensor fusion
\end{IEEEkeywords}

\footnotetext[1]{The authors contributed equally to this work}
\let\thefootnote\relax\footnotetext{This is the accepted version of the article published in IEEE Journal of Microwaves, DOI: \href{https://doi.org/10.1109/JMW.2026.3713325}{10.1109/JMW.2026.3713325}. \copyright~2026 The Authors. This work is licensed under a Creative Commons Attribution 4.0 License. For more information, see \url{https://creativecommons.org/licenses/by/4.0/}}

\section{Introduction}
In recent years, the reconstruction of dynamic targets using contactless sensors has gained significant attention, influencing research in many areas, including entertainment (e.g., computer games, AR/VR), autonomous agents, human-computer interaction, and medical diagnosis~\cite{yunus_2024, wang_2024}.

Among these, applications involving critical decisions based on complex movements, such as human gait analysis~\cite{seifert_2019, gambietz_2024} and clinical hand function assessments~\cite{amprimo_2024, wirth_2023, phutane_2021}, place a particularly high demand on fast and precise sensing techniques to ensure the reliability of such decisions.

Millimeter-wave (mmWave) multiple-input multiple-output (MIMO) radars offer a viable solution, providing spatial resolution beyond the capabilities of traditional monostatic antenna systems, and enabling the distinction of static and dynamic targets.
Moreover, radar systems can analyze motion via the Doppler effect, making them well-suited for dynamic environments, compared to other modalities such as conventional LiDAR or RGB-D cameras.
Thus, MmWave MIMO radar systems have been utilized for 3D human body reconstruction~\cite{chen_2022, chen_2023}, pose estimation \cite{lee_2023, engel_2025}, people tracking~\cite{zewge_2019}, and activity recognition~\cite{singh_2019}.

To achieve high-resolution three-dimensional imaging, MIMO radars typically employ a large number of transmitting (TX) and receiving (RX) antennas. 
\VWedit{With the ongoing move towards larger apertures~\cite{wirth_2024}, these systems increasingly operate in the sensor’s near field.}
Traditional \VWedit{near-field} radar imaging techniques, such as backprojection~\cite{wolf_1969, sherif_2021}, rely on these \VWedit{large and} dense antenna arrays to leverage the numerous TX-RX combinations for precise reconstruction;
however, \VWedit{the algorithmic complexity of the volumetric space, that is sampled in near-field backprojection, scales with the number of RX-TX antenna pairs, implying that precise 3D imaging capabilities of high-resolution antenna arrays} come \VWedit{along with} the expense of significant computational resources \VWedit{for reconstruction}. 
\VWedit{Moreover, in representative applications such as security screening~\cite{sherif_2021}, where frequency-stepped continuous-wave (FSCW) radars rely on time-division multiplexing (TDM), high‑fidelity 3D imaging via backprojection requires a high bandwidth and a substantial number of sequentially transmitted frequency tones. 
Consequently, this setting imposes high constraints on the attainable capture rate and, in turn, renders these sensors currently impractical for dynamic targets.
In this work, we address these limitations, aiming to meet the stringent demands for fast capture rates and computational efficiency alongside high-fidelity reconstruction for dynamic scenes.}

\VWedit{A line of prior work aims to reduce computation and capture time by utilizing sparse, low-cost antenna arrays.} 
Recent advancements in deep learning focus on smooth \VWedit[neural target]{reconstruction} representations that surpass the spatial resolution limitations of conventional signal processing techniques. 
One approach is to employ generative methods, such as implicit neural representations~\cite{farrell_2025} or conditional generative adversarial networks~\cite{guan_2020}, to recover the spatially resolved reflective properties of targets at super-resolution. 
Another line of work uses Neural Radiance Fields (NeRFs) as compact geometric representations to simulate novel views and synthesize raw frequency-space measurements~\cite{borts_2024} or range-Doppler maps~\cite{huang_2024}, while also incorporating additional modalities such as LiDAR and cameras~\cite{rafidashti_2025}.
Due to their data-driven learning processes, these approaches typically require a substantially high number of radar measurements, which currently limits their application primarily to autonomous driving scenarios. 
Moreover, the training process is computationally intensive.

\VWedit{Closer to our work, another line of research utilizes dense antenna arrays in combination with a reduced set of frequency tones. }
\VWedit{Among these approaches}, Bräunig et al.~\cite{braeunig_2023_fsk} introduced an imaging technique for large, densely-packed antenna arrays of high-resolution MIMO radars, which focuses on high-speed pose tracking of both static and dynamic human hands.
The proposed \textit{2FSK} method operates using just two neighboring frequencies, based on the principle of continuous-wave (CW) Frequency Shift Keying (FSK).
This approach significantly speeds up the reconstruction process \VWedit{by a factor of 1000 compared to} backprojection~\cite{braeunig_2023_fsk}, while ensuring rapid capture times due to the limited \VWedit{set of sequentially transmitted frequency tones}

A key limitation of the 2FSK method, however, is its assumption that the depth of the captured object is \VWedit[roughly]{approximately} known. Follow-up work~\cite{braeunig_2024} (\textit{3FSK}) aims to improve robustness by requiring more complex hardware configurations, that is, a larger signal bandwidth and three representative frequencies with specific frequency displacements, allowing for a less accurate depth prior.

The initial scalar depth prior, \VWedit{used by 2FSK and 3FSK, resembles} a plane or depth slice in \VWedit{in a discretized, volumetric} 3D space; \VWedit{hence, it renders} both methods suitable primarily for flat targets with limited depth extent, as shown for hands in~\cite{braeunig_2023_fsk, braeunig_2024}. 
Consequently, their applicability is restricted to a narrow range of target geometries and can lead to inaccuracies in \VWedit{environments, where depth is uncertain.}

\VWedit{We extend 2FSK to a multimodal setting (MM-2FSK) that preserves the two‑tone capture and reconstruction speed while supporting complex, spatially varying surface geometry. 
The key idea is to incorporate an absolute per-point surface-depth prior from a secondary modality into the 2FSK phase model, thereby enabling fine-grained relative depth estimation of that surface via the FSK principle. 
We revisit the conditions for unambiguous depth recovery with two tones and detail how to embed this external prior into the spatial sensor coordinate system and the signal model of the primary radio-frequency modality, while explicitly accounting for phase-range ambiguities and data mismatch between modalities.
Specifically, our method targets high‑resolution, near‑field mmWave FSCW MIMO imaging radars as primary modality, while we rely on a secondary sensor that supplies only the absolute surface prior. 
In principle, this prior is modality‑agnostic and can be sourced from any secondary sensor, including low-resolution measurements, e.g., from another mmWave sensor, when sensing characteristics within similar frequency bands are favored.
In practice, we demonstrate multimodal efficiency by using a secondary sensor from the optical spectrum, i.e. an RGB-D camera, due to its high spatial resolution.}

\begin{figure*}[!htbp]
	\centering
	\includegraphics[width=0.9\linewidth]{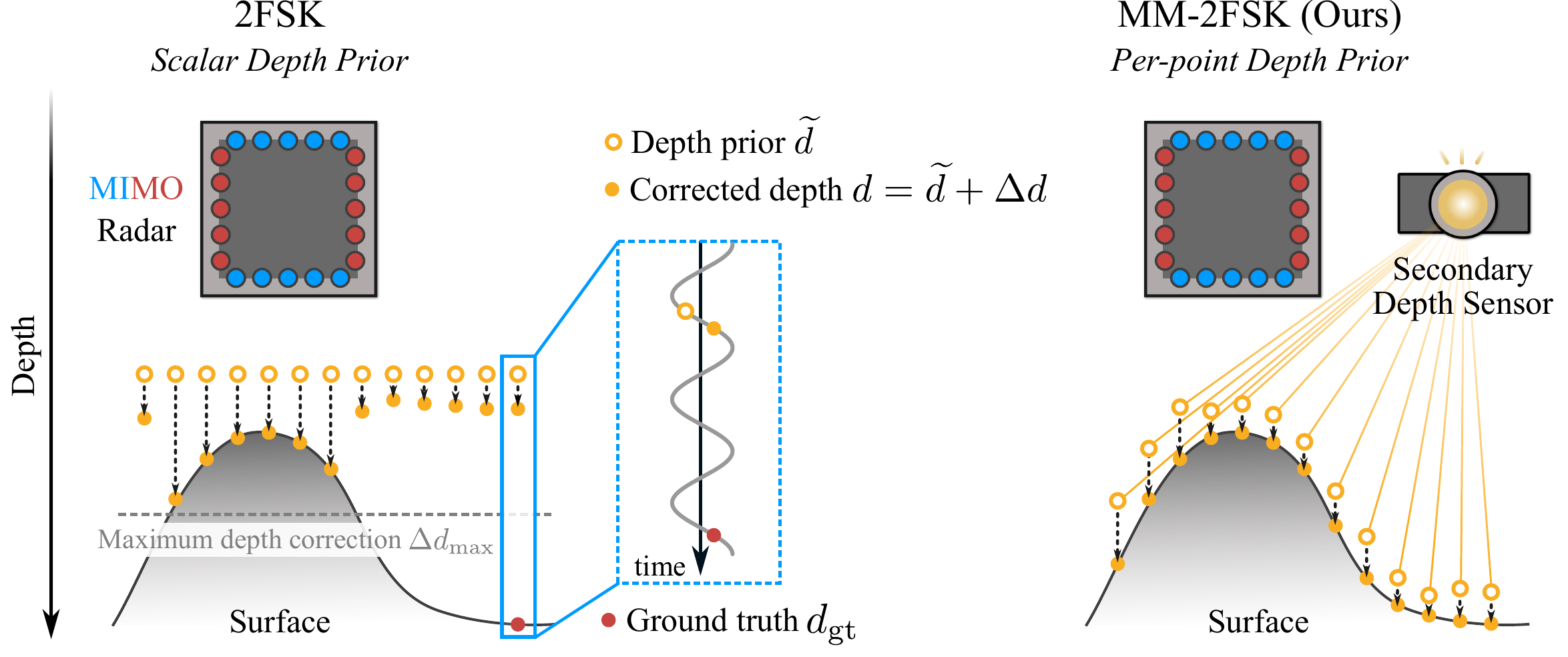} 
	\caption{In our work, we extend the 2FSK imaging principle to the multimodal domain (MM-2FSK). Given a unified scalar depth prior, $\widetilde{d}$, for each point, the 2FSK method iteratively refines the current estimate with a per-point depth correction factor, $\Delta d$, up to a limited extent, given by the maximum unambiguous depth correction, $d_{\text{max}}$. In contrast, our method receives per-point depth priors from a secondary depth sensor, without requiring knowledge about the target position, and is more robust towards targets of varying surface depth.}
	\label{fig:overview}
\end{figure*}

We provide a comprehensive evaluation based on a dataset~\cite{wirth_2024} that \VWedit{includes} ground-truth geometry of static objects, spatially aligned with real-world measurements from a mmWave high-resolution MIMO imaging radar \VWedit{operating in its near field}.
In this evaluation, we compare {MM-2FSK} against 2FSK, 3FSK, and traditional \VWedit{near-field} backprojection.
In addition, we investigate the influence of signal bandwidth beyond theoretical analysis and provide ablation studies focused on different frequency configurations.
\VWedit{Lastly, we demonstrate effective, robust reconstruction of moving targets in dynamic settings.}

In summary, our contributions are the following:
\begin{enumerate}
	\item A novel multimodal signal processing method that incorporates a mmWave FSCW MIMO radar along with \VWedit{a secondary, assistive modality.} \VWedit{In experiments, we selected secondary sensors from the optical frequency spectrum: an} active stereo camera, \VWedit{a time-of-flight} camera\VWedit{, and a multi-view stereo system}.
	\item A method for robust, high-speed radar imaging of arbitrary object \VWedit{geometries} without requiring additional knowledge about the capture environment, i.e. the object position and \VWedit{surface} depth variation.
	\item A comprehensive evaluation \VWedit{in static and dynamic settings}, \VWedit{which includes an} ablation study over different frequency configurations and comparison to state-of-the-art radar imaging methods, i.e. 2FSK, 3FSK, and backprojection.
\end{enumerate}

\section{Frequency Shift Keying for MIMO Radar Imaging}
In the following section, we first address the theoretical foundations of the 2FSK imaging principle of Bräunig et al.~\cite{braeunig_2023_fsk}.
Subsequently, we derive our MM-2FSK method from this principle.
The key differences between the two algorithms are highlighted in~\autoref{fig:overview}.

Both methods are designed for high-resolution MIMO imaging radars that utilize multiple \VWedit{TX-RX} antenna pairs for imaging. 
For simplicity, we omit the repetitive calculations over multiple antenna pairs\,--\, typically performed for \VWedit{near-field} backprojection and related imaging methods\,--\,and present exemplary equations using just one TX-RX antenna pair. 
For a more detailed derivation of the equations, we refer to~\cite{braeunig_2023_fsk}.

\subsection{Frequency Shift Keying with Two Neighboring Frequencies (2FSK)}
The 2FSK approach uses two transmitted signals of discrete neighboring frequencies, $f_1$ and $f_2$.
\VWedit{As is typical in MIMO antenna configurations, the corresponding baseband signals of these two frequencies are first transmitted from a TX antenna, positioned at $\boldsymbol{r}_{\text{TX}} \in \mathbb{R}^3$.
These signals then propagate through space until they reflect off the first point target located at ${\boldsymbol{p} \in \mathbb{R}^3}$. Finally, the signal reflections are received by all RX antennas, with each positioned at a different location ${\boldsymbol{r}_{\text{RX}} \in \mathbb{R}^3}$.}
After signal demodulation, the baseband signals, $s_i$ with $i\in\{1,2\}$, can be expressed in analytic notation as follows:
\begin{equation}
	s_{i} = A_{i} \text{exp}\left( -j 2\pi f_{i} \frac{\rho}{\text{c}} + \phi_c \right)\, ,
\end{equation}
where $A_i$ is the amplitude, $\text{c}$ is the speed of light, and $\phi_c$ is a constant phase offset.
The traveled round-trip distance to $\boldsymbol{p}$, defined as $\rho = \lVert \boldsymbol{r}_{\text{TX}} - \boldsymbol{p} \rVert_2 + \lVert \boldsymbol{r}_{\text{RX}} - \boldsymbol{p} \rVert_2$, relates to the target depth $d$ by $\rho = 2d$, assuming far-field conditions where depth approximates range.

Given a set of candidate point target locations ${\widetilde{\boldsymbol{p}} \in \boldsymbol{\mathcal{P}} = \{(x,y,\widetilde{d})\}}$ with a corresponding scalar depth prior $\widetilde{d}$, two signal hypotheses, $w_{1}, w_{2}$, are computed from the round-trip distance $\widetilde{\rho}$ between a TX-RX antenna pair and $\widetilde{\boldsymbol{p}}$:
\begin{equation}
w_{i}(\widetilde{\rho}) = \text{exp}\left( -j 2\pi f_{i} \frac{\widetilde{\rho}}{\text{c}} \right)\, .
\end{equation}
The hypotheses are correlated with the baseband signals as follows:
\begin{equation}
	c_{i}(\widetilde{\rho}) = s_{i} w_{i}(\widetilde{\rho})\ast = \text{exp}\left( -j 2\pi f_{i} \frac{(\rho- \widetilde{\rho})}{\text{c}} \right) \, ,
\label{eq:backproj}
\end{equation}
where $*$ denotes the complex conjugate.
The resulting complex signal contains a residual phase $\Delta \varphi_{i}$ that is proportional to a correction factor for distance, $\Delta \rho = (\rho - \widehat{\rho}) = 2\Delta d$, and, correspondingly, \VWedit{a factor for} depth $\Delta d$:
\begin{align}
	 \label{eq:phase_to_depth}
	 2\pi f_{i} \frac{\Delta \rho}{\text{c}} &= 2\pi f_{i} \frac{2 \Delta d}{\text{c}} = \Delta \varphi_{i} \\
	 \Leftrightarrow \Delta d &= \frac{\text{c}\Delta \varphi_{i}}{4\pi f_{i}}\, ,
	 \label{eq:delta_depth}
\end{align}
With this information, each per-point depth estimate $d$ can be refined as follows:
\begin{equation}
	d = \widetilde{d} + \Delta d\, .
	\label{eq:depth_correction}
\end{equation}

Computing the phase \VWedit{$\Delta \varphi_i$} involves inverse trigonometric functions to determine the angle of the residual complex phasor $c_{i}(\widetilde{\rho})$.
Due to the $2\pi$ phase ambiguities arising from its repetitive nature, these angles are typically restricted to the first period of the residual phasor.
Consequently, the maximum correction factor for depth in one direction can be derived from \autoref{eq:delta_depth} by assuming $\Delta \varphi_i$ approaches $2\pi$, which yields $\text{c}/(2 \cdot f_{i})$.

\begin{figure}[!htbp]
	\centering
	\includegraphics[width=0.9\linewidth]{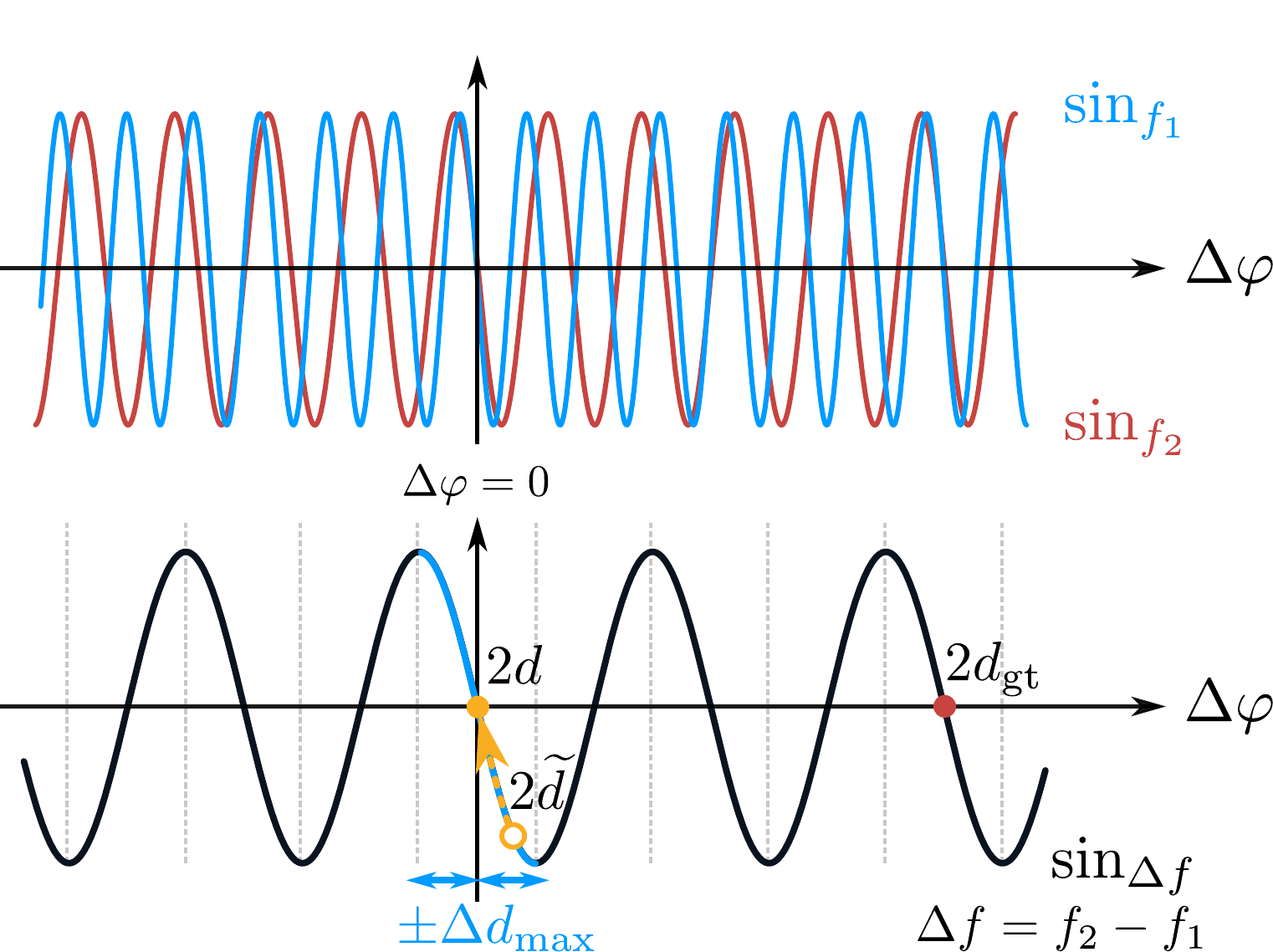} 
	\caption{
		Simplified visualization of the 2FSK depth correction process, where complex, analytic signals are schemed as periodic, real-valued sine waves.	
		The 2FSK principle computes the depth correction $\Delta d$ based on the residual of the phase, $\Delta\varphi$, that remains after correlating the two single-frequency signals with a signal hypothesis, constructed with the depth prior $\widetilde{d}$.
		The \textit{top row} depicts the two residual signals at frequencies $f_1$ and $f_2$, respectively.
		Using these, a complex differential signal with frequency $\Delta f$ is calculated, as depicted in the \textit{bottom row}.
		This differential signal is used to adjust the current depth guess and is constrained by the maximum unambiguous depth correction, $\pm d_{\textit{max}}$.
		The correction factor is centered around the zero-crossing of the signal within the first period\,--\,or here, half of the period, due to signal simplification\,--\, pointing into the direction where the residual phase yields zero.
		Due to the $2\pi$-periodicity of the continuous signal, the residual phase corresponding to the ground truth depth, $d_{\text{max}}$ may lie within a different signal period, resulting in the depth correction not producing the intended outcome.}
	\label{fig:unambiguous_range}
\end{figure}

High-resolution MIMO imaging radars typically operate in the GHz to THz range~\cite{sherif_2021}, \VWedit{implying that} this maximum correction factor can be small\VWedit{, i.e., within millimeters}. 
Thus, Bräunig et al.~\cite{braeunig_2023_fsk} introduced the concept of calculating a differential complex phasor from the two single-frequency residual phasors based on~\autoref{eq:backproj} and \autoref{eq:phase_to_depth} as follows:
\begin{equation}
	\label{eq:diff_signal}
	c_{\Delta f}(\widetilde{\rho}) = c_{2}(\widetilde{\rho})c_{1}(\widetilde{\rho})\ast = \text{exp}\left(-j 2\pi \Delta f \frac{2 \Delta d}{\text{c}} \right)\, .
\end{equation}
The complex phasor of frequency difference $\Delta f = f_2 - f_1$ resembles a signal of considerably lower frequency (cf.~\autoref{fig:unambiguous_range}), allowing a depth correction $\Delta d$ to be within the so-called \textit{maximum unambiguous depth correction} $\Delta d_{\text{max}}$, which now solely depends on the signal bandwidth:
\begin{equation}
\label{eq:depth_from_diff}
\Delta d = \frac{\text{c}\Delta \varphi_{\Delta f}}{4 \pi \Delta f} \Rightarrow \Delta d_{\text{max}} = \frac{\text{c}}{2\cdot 2\Delta f}\, .
\end{equation}
It is noteworthy that the right side corresponds to Equation~8 from~\cite{braeunig_2023_fsk}, additionally divided by a factor of 2 since we consider depth correction in both directions, yielding ${\Delta d \in [-\Delta d_{\text{max}}, +\Delta d_{\text{max}}]}$.

We illustrate the intuition behind the depth correction in \autoref{fig:unambiguous_range}, where we simplify the illustration of an analytic complex signal to a simple sine wave.
As the depth correction is limited to the first period of the residual phasor $c_{\Delta f}$, the 2FSK algorithm does not necessarily converge to the ground-truth depth $d_{\text{max}}$ of each point target, which may lie outside this period. 
Consequently, depth correction can fail and may even inadvertently adjust prior depth estimates in the wrong direction.

\subsection{Multimodal Frequency Shift Keying (MM-2FSK)}
\VWedit{Our method extends high‑speed mmWave radar imaging to variable surface geometry by utilizing per-point depth priors from a secondary modality, ensuring the precision meets the unambiguous range-phase requirement for depth correction using a FSCW signal modulation.
We detail how to embed this prior into the near‑field MIMO coordinate and signal processing model while mitigating errors induced by sensor data mismatches in sampling and resolution. 

Furthermore, we target the method's practical operation in dynamic scenes, where concurrent reconstruction and capture is desirable, by providing an efficient GPU implementation that integrates readily with matched‑filter–based imaging, including backprojection, 2FSK, and MM-2FSK.}

\subsubsection{Algorithm}
\VWedit{
Unlike 2FSK, which relies on a single scalar depth prior, MM-2FSK leverages per-pixel depth from a secondary modality; in the following, we describe the algorithm using a generalized, modality-agnostic formulation.
We assume that the secondary sensor is spatially calibrated with the MIMO imaging radar. 
Such spatial calibration can be obtained, for instance, with target-based procedures suitable for near-field operation, e.g., by employing symmetrically arranged spherical calibration targets composed of metallic and styrofoam materials~\cite{wirth_calibration}.
The resulting extrinsic parameters allow transforming the 3D points from the secondary modality into the coordinate system of the primary modality, i.e. the imaging radar.
}

\VWedit{We assume the secondary modality provides a depth map $\boldsymbol{D}_s \in \mathbb{R}^{H \times W}$, where each pixel coordinate $(u,v)$ with $u\in[0,W-1]$ and $v\in[0,H-1]$ is associated with a metric depth scalar $d$.
We adopt a generalized ray model to transform each per-pixel depth measurement $(u,v,d)$ into its respective 3D point, while considering different image projection models.
The 3D point is described by a ray in the sensor's coordinate system that is defined by an origin, $\boldsymbol{o}(u,v)\in \mathbb{R}^3$ and a direction $\boldsymbol{\omega}(u,v)\in \mathbb{R}^3$, that is scaled by the depth $d$, such that:
\begin{equation}
	\boldsymbol{r}(u,v,d) = \boldsymbol{o}(u,v) + d\cdot\boldsymbol{\omega}(u,v)\; .
\end{equation}

For range sensing modalities, where $\boldsymbol{o}(u,v)$ and $\boldsymbol{\omega}(u,v)$ are directly provided by the sensor, this formulation applies in a straightforward manner. 
We further specialize the model to two common projection types used in high-resolution depth imaging: perspective and orthographic projection.

Perspective projection is typically associated with sensors based on pixel-array imaging hardware, such as standard cameras.
Here, the ray that points to the corresponding 3D point of a pixel is typically derived from the intrinsic calibration parameters of a pinhole camera model:
\begin{align}
\boldsymbol{o}(u,v) &= (0,0,0)^T  \\
\boldsymbol{\omega}(u,v) &= 
\left(
\begin{array}{ccc}
f_u & 0 & c_u \\
0 & f_v & c_v \\
0 & 0 & 1
\end{array}\right)^{-1}
\begin{pmatrix}
u \\ v \\ 1
\end{pmatrix}\;,
\label{eq:project}
\end{align}
where $f_u, f_v$ are the focal lengths and $c_u, c_v$ are the principal point offsets.

Another class of sensors is based on orthographic projection models. 
Typical representatives are imaging radar systems that reconstruct the scene via near-field backprojection, followed by Maximum Intensity Projection (MIP)~\cite{wirth_2024} along the depth axis of a Cartesian coordinate system defining the spatial extent and location of the reconstruction volume around the target of interest.
For orthographic projection, the generalized ray model maps the origin to metric coordinates along the lateral dimensions, and the ray direction reduces to an identity vector:
\begin{align}
\boldsymbol{o}(u,v) &= \left(
\begin{array}{cccc}
\frac{(N_x - 1)}{r-l} & 0 & 0 & \frac{(N_x - 1) \cdot l}{l-r} \\
0 & \frac{(N_y - 1)}{b-t} & 0 & \frac{(N_y - 1) \cdot b}{t-b} \\
0 & 0 & 1 & 0
\end{array}\right)^{-1}
\begin{pmatrix}
u \\ v \\ 0 \\ 1
\end{pmatrix}\;, \\
\boldsymbol{\omega}(u,v) &= (0,0,1)^T\,
\end{align}
where $(l,r)$ and $(t,b)$ are the spatial min-max extents of the reconstruction volume along the Cartesian x- and y-axis, respectively, and $N_x, N_y$ denote the respective number of discretized voxels within these extents.
}

\VWedit{Moreover}, a depth image from a secondary \VWedit{modality} may contain invalid pixels, e.g., due to the sensitivity to environmental \VWedit{factors}; such pixels are simply skipped.

\begin{figure}[!htbp]
	\centering
	\includegraphics[width=\linewidth]{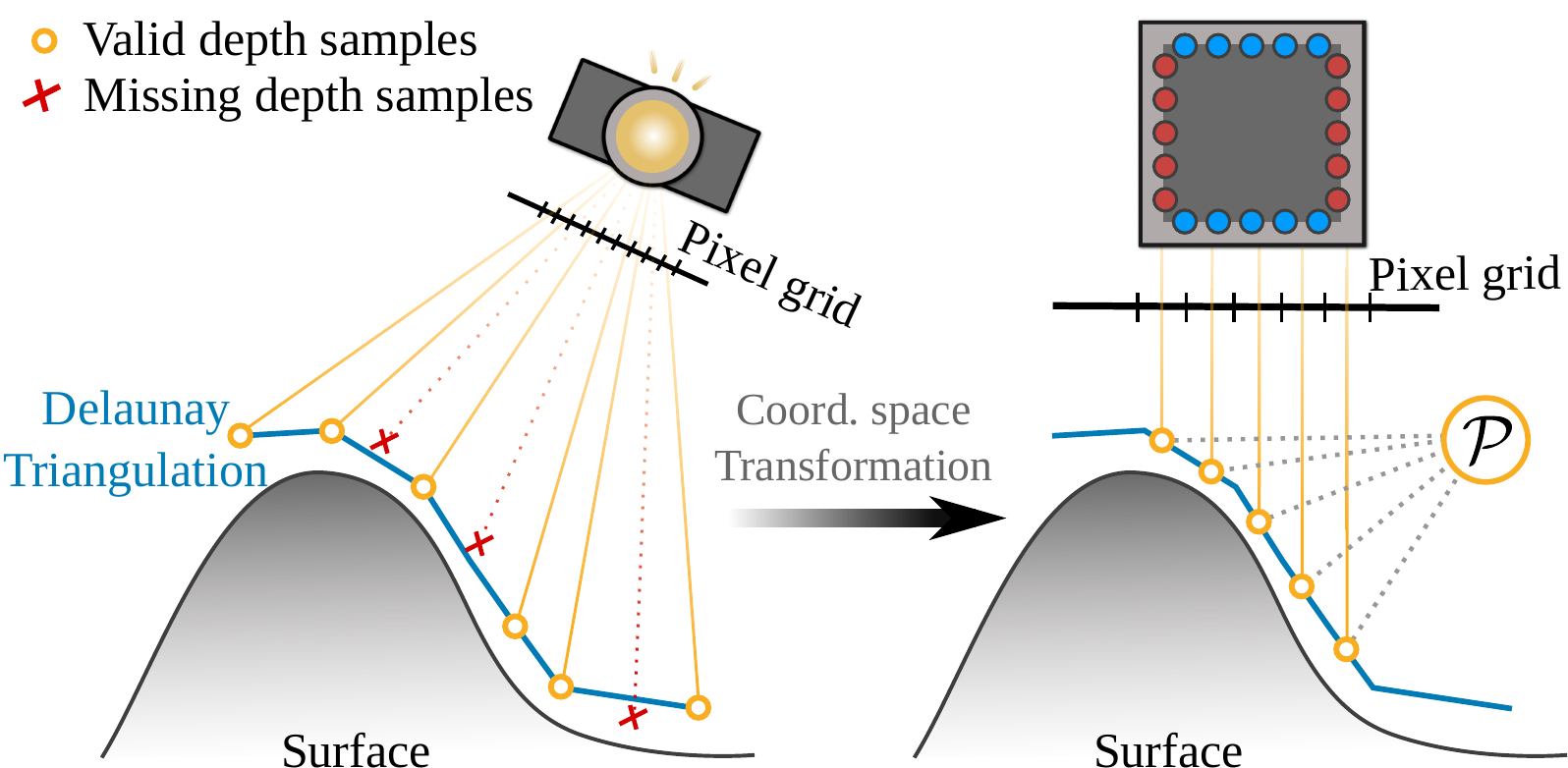} 
	\caption{
		Visualization of the depth prior generation: We first create a closed triangle mesh using Delaunay triangulation on 2D pixels corresponding to valid 3D point samples from the optical depth sensor; illustrated here in 2D as blue line sets. The mesh is then transformed into the radar's coordinate space and re-sampled via rasterization on the radar pixel grid to generate the candidate point set $\mathcal{P}$.}
	\label{fig:depth_prior}
\end{figure}

To generate a \VWedit{spatially coherent} depth prior for radar imaging, \VWedit{we convert} the resulting \VWedit{3D} point cloud \VWedit{$\boldsymbol{P}_s \in \mathbb{R}^{N\times3}$} to a closed triangle mesh. %
To this end, we triangulate the point cloud in 2D using Delaunay triangulation~\cite{delaunay_1934} \VWedit{based on the corresponding pixel coordinates $(u,v)$}.
This triangulation computes the 2D convex hull of $\boldsymbol{P}_s$, effectively filling in depth gaps and preventing surface holes.
A visualization of such triangulation is given in~\autoref{fig:depth_prior}.
Next, we use the extrinsic parameters obtained from spatial calibration~\cite{wirth_calibration} to transform the triangulated point cloud into the radar's coordinate space:
\begin{equation}
\boldsymbol{P}^r_s
= 
[\boldsymbol{R} \mid \boldsymbol{t}]
\boldsymbol{P}_s\, .
\label{eq:warp}
\end{equation}
The extrinsic parameters consist of a rotation $\boldsymbol{R}\in \mathbb{R}^{3\times3}$ and translation $\boldsymbol{t}\in \mathbb{R}^{3}$.

We then construct the set of candidate point target locations $\boldsymbol{\mathcal{P}}$. 
Since the final output of high-resolution MIMO radars is typically an image, we compute $\boldsymbol{\mathcal{P}}$ by sampling a user-defined ${H' \times W'}$ pixel grid of Cartesian 2D coordinates $(x,y)$, centered around the antenna aperture.
Given the \VWedit{orthographic} mapping of spatial coordinates to radar image pixels, we rasterize the triangulated point cloud $\boldsymbol{P}_s^r$.

Specifically, we render a depth map with barycentrically interpolated depth values, based on the triangle topology, to yield $\boldsymbol{\mathcal{P}}\in \mathbb{R}^{H'W'\times 3}$.
This essentially becomes our set of point candidates with \textit{per-point} depth prior $\widetilde{d}$, as illustrated in~\autoref{fig:depth_prior}.
Finally, we proceed with the depth correction in analogy to \autoref{eq:depth_correction} of the 2FSK method.

Note that in contrast to 2FSK and 3FSK our depth prior is not constant and can thus represent non-flat shapes with larger depth range.
As long as the noise parameters of the \VWedit{secondary modality} and any spatial calibration errors remain within the maximum unambiguous depth correction factor, the final depth estimates of the MM-2FSK method are expected to be close to the ground truth.
\newcommand\mycommfont[1]{\footnotesize\ttfamily\textcolor{blue}{#1}}
\SetCommentSty{mycommfont}
\begin{algorithm}
	\caption{(MM-)2FSK for one CUDA thread}
	\label{alg:fsk}
	\KwIn{Thread ID $i\in [0, 31]$, warp ID $j \in \mathbb{N}_0$, baseband signals $\boldsymbol{S}_1, \boldsymbol{S}_2 \in \mathbb{C}^{T \times R}$, point candidates $\boldsymbol{\mathcal{P}} \in \mathbb{R}^{H'W'\times 3}$, antenna positions $\boldsymbol{R}_{\text{TX}} \in \mathbb{R}^{T \times 3}$ and $\boldsymbol{R}_{\text{RX}} \in \mathbb{R}^{R \times 3}$}
	\KwData{Shared memory buffer $\boldsymbol{D}_{\boldsymbol{r}_{\text{TX}}} \in \mathbb{R}^T$, $\boldsymbol{D}_{\boldsymbol{r}_{\text{RX}}} \in \mathbb{R}^R$}
	\KwOut{Correlations $\boldsymbol{C}_1 \in \mathbb{C}^{H' W'}$, $\boldsymbol{C}_2 \in \mathbb{C}^{H'W'}$}
	$\boldsymbol{p} \leftarrow \boldsymbol{\mathcal{P}}[j]$\;
	\For{$k \leftarrow 0,T/32$}{
		$\boldsymbol{D}_{\boldsymbol{r}_{\text{TX}}}[i + 32 \cdot k] \leftarrow \lVert \boldsymbol{R}_{\text{TX}}[i + 32 \cdot k] - \boldsymbol{p} \rVert_2$\;
	}
	\For{$k \leftarrow 0,R/32$}{
		$\boldsymbol{D}_{\boldsymbol{r}_{\text{RX}}}[i + 32 \cdot k] \leftarrow \lVert \boldsymbol{p} - \boldsymbol{R}_{\text{RX}}[i + 32 \cdot k] \rVert_2$\;
	}
	$\bar{c}_1, \bar{c}_2 \leftarrow 0$\;
	\For{$k \leftarrow 0,(T\cdot R)/32$}{
		$\widetilde{\rho} \leftarrow \boldsymbol{D}_{\boldsymbol{r}_{\text{TX}}}[i + 32 \cdot k] + \boldsymbol{D}_{\boldsymbol{r}_{\text{RX}}}[i + 32 \cdot k]$\;
		$\bar{c}_1\leftarrow \bar{c}_1 + c_1(\widetilde{\rho})$\tcp*[r]{ \autoref{eq:backproj} for $f_1$} 
		$\bar{c}_2\leftarrow \bar{c}_2 + c_2(\widetilde{\rho})$\tcp*[r]{ \autoref{eq:backproj} for $f_2$} 
		
	}
	$\boldsymbol{C}_1[j] \leftarrow \text{warp\_reduce\_sum}(\bar{c}_1) / (T \cdot R)$\;
	$\boldsymbol{C}_2[j] \leftarrow \text{warp\_reduce\_sum}(\bar{c}_2) / (T \cdot R)$\;
\end{algorithm}
\subsubsection{Efficient Implementation}
Our method utilizes the single-instruction multiple-threads (SIMT) instructions of graphics processing units (GPUs), implemented by using NVIDIA CUDA as domain specific language.
In this section, we put emphasis on the CUDA implementation of the baseband signal correlation kernel, which is commonly the runtime bottleneck in reconstruction using high-resolution imaging radars.

In MIMO radar imaging algorithms~\cite{braeunig_2023_fsk, sherif_2021}, determining the spatial position of a point target typically requires performing correlation across the entire antenna aperture configuration. Specifically, for each candidate point target {$\boldsymbol{p} \in \boldsymbol{\mathcal{P}}$}, the residual phasors from~\autoref{eq:backproj} are averaged over multiple TX-RX antenna positions. 
To optimize performance on the GPU, we parallelize the iterations over point targets and antenna pairs, utilizing the shared memory features of the GPU architecture.

NVIDIA graphics cards consist of SIMT units called \textit{warps}, which include 32 parallel threads grouped into blocks that share fast-access memory. 
In our GPU kernel, as shown in~\autoref{alg:fsk}, each point $\boldsymbol{p}$ is processed by an entire warp, distributing the computations over multiple TX-RX antenna pairs. 
Given the known $T \times R$ MIMO antenna architecture, each thread pre-computes a subset of one-directional antenna paths from any TX antenna to $\boldsymbol{p}$ and from $\boldsymbol{p}$ to any RX antenna, stored in \( \boldsymbol{D}_{\boldsymbol{r}_{\text{TX}}} \) and \( \boldsymbol{D}_{\boldsymbol{r}_{\text{RX}}} \).

By sharing memory within the warp, each thread computes a partial summation of the residual phasors over all possible ${(T\cdot R)/32}$ TX-RX combinations. 
Finally, we utilize CUDA warp functions for summing the residual phasors in \texttt{warp\_reduce\_sum} and then average the result. 
The depth correction step, calculated from~\autoref{eq:diff_signal}, is then performed after the kernel, as we use the two intermediate residual phasor sets, $\boldsymbol{C}_1$ and $\boldsymbol{C}_2$, for additional depth filtering, as discussed in the next section.

\section{Experimental Setup}
In the following sections, we describe the measurement setup derived from the MAROON dataset~\cite{wirth_2024}, along with the implementation details of the algorithms that we use in our evaluation. 
\VWedit{We compare our method against:} 
\begin{enumerate}
	\item Near-field backprojection (\VWedit{Eq.~1 in}~\cite{sherif_2021}), \VWedit{followed by MIP (Eq.~8 in~\cite{wirth_2024}) along the depth axis}.
	\item 2FSK \VWedit{(Eq.~\ref{eq:depth_correction}--\ref{eq:depth_from_diff} in this work)}.
	\item 3FSK \VWedit{(Fig.~2 in~\cite{braeunig_2024})}.
\end{enumerate}
Similar to ours, the 3FSK method extends the 2FSK approach by utilizing three representative frequencies with specific frequency displacements, resulting in three frequency differences. 
This allows for depth correction to be performed twice: first, by using the initial scalar depth value with a low frequency difference and, second, by utilizing the per-point corrected depth prior with two high frequency differences.

\subsection{\VWedit{Capture Setup}}
\begin{figure}[!htbp]
	\centering
	\includegraphics[width=0.7\linewidth]{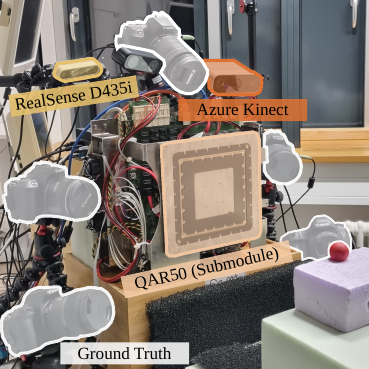} 
	\caption{
		\VWedit{Sensor configuration from the MAROON dataset used in our experiments. For static-scene evaluation with MM-2FSK, we utilize per-point depth priors from an Intel RealSense D435i (active stereo) camera, selected for its low-noise characteristics~\cite{wirth_2024}. For evaluating dynamic scenes, we pair the QAR50 radar submodule with a Microsoft Azure Kinect (optical time-of-flight) camera due to its comparatively simple time-synchronization support.}}
	\label{fig:setup}
\end{figure}

We validate the proposed method using the \VWedit{sensor configuration from the} MAROON dataset~\cite{wirth_2024}. 
\VWedit{The} measurements \VWedit{were} collected from a high-resolution MIMO radar, which \VWedit{is depicted in~\autoref{fig:setup}, where it is arranged alongside} three spatially calibrated depth cameras and a ground-truth \VWedit{multi-view stereo (MVS)} measurement system \VWedit{composed of five digital single-lens reflex (DSLR) cameras}. 

The QAR50 MIMO radar submodule features a \VWedit{${13.8\times13.8~\text{cm}}$ square-shaped} aperture consisting of $94 \times 94$ TX-RX antenna pairs. \VWedit{The sensor operates in its near field. It} employs a \VWedit{FSCW} signal modulation \VWedit{with a maximum of} 128 \VWedit{frequency tones} \VWedit{ranging} from 72 to 82~GHz. \VWedit{Further sensor specifications are listed in Table~1 of Wirth et al.~\cite{wirth_2024}}.

\VWedit{To evaluate static scenes of varying geometry, we rely on the sensor captures from MAROON that} comprise real sensor measurements of 45 distinct household and construction objects. 
\VWedit{Each target is positioned such that its geometric center coincides with the aperture center of the antenna array. 
To align the centers along the height dimension, the targets are supported by styrofoam boards (cf.~\autoref{fig:setup}).}

While the target objects were captured at various distances~\cite{wirth_2024}, we focus on the object measurements taken at approximately 30~cm \VWedit{radial distance} from the MIMO radar.
To assess the imaging accuracy of the proposed method, we utilize ground-truth measurements of the target objects, obtained from \VWedit{the MVS system}. 

\VWedit{For evaluating dynamic scenes, we manually captured two additional sequences of radial hand motion, using a subset of the sensor arrangement given in~\autoref{fig:setup}, where we only utilize the QAR50 and the Azure Kinect time-of-flight sensor due to the availability of time-synchronization support of both sensors. 
In these measurements, the hand pose is held constant and the arm position is varied back and forth between approximately 20~cm and 60~cm radial distance to the imaging radar.
In the absence of reliable surface ground truth\;--\;since dynamic scenes are challenging to reconstruct with traditional MVS\;--\;we perform a relative comparison against high-resolution backprojection, utilizing the full signal bandwidth and 128 frequency steps.}

Depending on the experiment, we \VWedit{utilize different secondary sensors for MM-2FSK: for static scenes}, we \VWedit{use} either the ground-truth system or the \VWedit{noise-resilient} {RealSense D435i} active stereo depth camera.
\VWedit{For dynamic scenes requiring inter-sensor time synchronization, we leverage the integrated timestamp mechanisms of the Azure Kinect camera.
Prior to data capture, we determine the time offset between relative Azure Kinect timestamps and the QAR50 timestamps by using the motion trajectory of a metal sphere pendulum as temporal calibration target. 
For evaluation, the modality-wise captures are then paired by nearest-neighbor timestamp matching.
We note that typical temporal data mismatch is expected even after calibration, caused by varying capture latency and the different frame rates of the optical sensor (30~fps) and the radar ($\approx$ 70~fps).
}

\subsection{Implementation Details}
\label{sec:implementation_details}

\VWedit{The MIMO imaging radar measurements provide} raw phasor data in the form of a $94\times94\times128$ complex tensor, which we reduce to include only the two or three relevant frequencies, resulting in a $94\times94\times2$ or $94\times94\times3$ tensor, respectively, depending on the radar imaging method.

For backprojection \VWedit{of static scenes}, we sample points within a ${30\times30\times 20}$~cm volume centered around the object, yielding a voxel grid of dimensions \VWedit{$N_x = N_y = 301$ and $N_z = 201$}.
This grid is then projected to a \VWedit{$H' \times W'= N_x \times N_y$} depth map using Maximum Intensity Projection \VWedit{along the depth axis (z-axis)}.

\VWedit{For backprojection of dynamic scenes, we sample points within a reconstruction volume of $l = b = -15~\text{cm}$, $r = t = 15~\text{cm}$, $n = 20~\text{cm}$, and $f = 60~\text{cm}$, where the pairs $(l,b)$, $(t,b)$, $(n,f)$ are the min-max extents along the Cartesian xyz-axes, respectively. 
The voxel density is kept at $N_x = N_y = 301$, consistent with the static case, while $N_z = 401$ is increased, effectively yielding twice as many voxels to accommodate the increased z-extent that is necessary to construct the radial hand motion.
}

For the 2FSK, 3FSK, and MM-2FSK methods \VWedit{in both static and dynamic scenarios}, we correspondingly reconstruct the \VWedit{$H' \times W'$} depth map directly.
To investigate realistic scenarios, where the object placement is not trivial to assess, we use a 2FSK/3FSK depth prior of $\widetilde{d} = 40$~cm, which means, \VWedit{for the static scenario} we expect a depth correction factor of $\Delta d \approx 10$~cm based on the ground truth at approximately 30~cm depth.

To filter out clutter and noise, we employ a depth filtering threshold across all four imaging methods, applied to the magnitude of the residual complex phasor after spatially resolving the signal.
Specifically, we use the CUDA kernel listed in~\autoref{alg:fsk} to average the residual phasors across all TX-RX antenna and frequency combinations, then compute the magnitude $V$ of the resulting mean phasor.
Finally, we keep the depth values with \VWedit{with respect to a decibel threshold, $20 \cdot \text{log}_{10}(\frac{V}{V_{\text{max}}}) \geq - 14$}~dB, relative to the maximum \VWedit{$V_{\text{max}}$} \VWedit{per object reconstruction}.

In terms of \VWedit{average} runtime, our implementation for depth estimation with two-frequency backprojection takes approximately 1430~ms, when using an NVIDIA GeForce RTX 3080 graphics card (10GB VRAM) and an Intel Xeon W-1390P (3.50 GHz) processor. 
3FSK achieves a runtime of about \VWedit{70}~ms and the (MM-)2FSK methods achieve a runtime of about \VWedit{40}~ms. %

\section{Evaluation}
\label{sec:evaluation}
\VWedit{This section} presents \VWedit{three} key experiments: first, an ablation study to explore the accuracy of MM-2FSK while varying the frequency differences. 
Second, a \VWedit{static-scene} comparison of our method against \VWedit{the state of the art, i.e.} 2FSK~\cite{braeunig_2023_fsk} and 3FSK~\cite{braeunig_2024} approaches, and traditional backprojection (BP)~\cite{sherif_2021, wolf_1969}.
\VWedit{Third, an extension of this state-of-the-art comparison to a dynamic radial hand motion setting.}
\VWedit{We perform evaluation using} six representative frequency configurations, which are listed in \autoref{table:2fsk_frequencies}. 

For {(MM-)2FSK} and BP, we use the terminology \textit{2FSK$\Delta f$}, e.g., {FSK$\Delta$0.5} to denote a configuration with 0.5~GHz frequency difference.
\VWedit{Furthermore, we denote the configuration of the lowest and highest frequency differences of 3FSK as {3FSK($\Delta f_{\text{min}}$, $\Delta f_{\text{max}}$)}.}

\begin{table}[!htbp]
	\caption{Frequency configurations, utilized for all subsequent experiments.}
	\begin{tabularx}{\linewidth}{@{}YYYY@{}} 
		\toprule 
		$\Delta f$ (GHz) & $f_1$ (GHz)  & $f_2$ (GHz) & $\Delta d_{\max}$ (cm)  \\ 
		\midrule
		$\approx$ 0.5 & 81.45 & 82.00  & 13.60 \\
		$\approx$ 1.0 & 80.98 & 82.00 & 7.32 \\
		$\approx$ 2.0 & 79.95 & 82.00 & 3.66 \\
		$\approx$ 4.0 & 77.91 & 82.00 & 1.83 \\
		$\approx$ 8.0 & 73.97 & 82.00 & 0.93 \\
		$\approx$ 10.0 & 72.00 & 82.00 & 0.75 \\
		\bottomrule
	\end{tabularx} 
	\label{table:2fsk_frequencies}
\end{table}

\subsection{Metrics}
We follow a similar evaluation procedure as outlined in the MAROON dataset, utilizing the corresponding metrics: the one-directional Chamfer distance and the projective error. 
Interested readers are referred to~\cite{wirth_2024} for a detailed discussion on the interpretation of these metrics.

The one-directional Chamfer distance quantifies the mean point-wise euclidean norm between point cloud ${\boldsymbol{P}_r \in \mathbb{R}^{N \times 3}}$, and point cloud $\boldsymbol{P}_{\text{gt}} \in \mathbb{R}^{M \times 3}$:
\begin{equation}
\text{C} = \frac{1}{N}\sum_{\boldsymbol{p}_r \in \boldsymbol{P}_r} \underset{\boldsymbol{p}_{\text{gt}} \in \boldsymbol{P}_{\text{gt}}}{\text{min}} \lVert \boldsymbol{p}_r - \boldsymbol{p}_{\text{gt}} \lVert_2\;.
\end{equation}
To compute this, we transform the radar depth maps back into point cloud representation, where we compare them against the re-sampled ground-truth object reconstruction of similar point density (cf.~\cite{wirth_2024}).
The Chamfer distance is measured in both directions: from the ground-truth point cloud to the radar reconstruction, denoted as \mone, and vice versa, denoted as \mtwo.

The projective error is calculated on the respective depth maps, $\boldsymbol{D}_r$ and $\boldsymbol{D}_\text{gt}$, with the ground-truth depth map obtained by rasterizing the point cloud with respect to the radar pixel grid:
\begin{equation}
\text{P} = \frac{1}{H' \cdot W'}\sum_{u=0}^{H'-1} \sum_{v=0}^{W'-1} \lvert \boldsymbol{D}_r(u,v) - \boldsymbol{D}_\text{gt}(u,v) \rvert\;.
\end{equation} 
Following the methodology of~\cite{wirth_2024}, we measure the projective error on the masked object depth maps, once with and without performing additional mask erosion to mitigate silhouette artifacts; we denote the resulting metrics as {\mthree} and {\mfour}, respectively.

\subsection{Ablation with Respect to Frequency Differences} 
\begin{table}[!htbp]
	\caption{Ablation study of the MM-2FSK method, \VWedit{using depth priors from MVS in combination} with different frequency configurations. All metrics are given in centimeters and are averaged over all \VWedit{static} objects at 30~cm distance. The \best{best} results per metric are highlighted.}
	\begin{tabularx}{\linewidth}{@{}cYYYY@{}} 
		\toprule 
		
		& { \mone } & { \mtwo }  & { \mthree }  & { \mfour } \\ 
		\midrule
		MM-2FSK$\Delta$0.5 & 0.72 & 2.14 & 2.15 & 1.69 \\
		MM-2FSK$\Delta$1.0 & 0.74 & 1.26 & 1.74 & 1.44 \\
		MM-2FSK$\Delta$2.0 & 0.57 & 0.60 & 0.77 & 0.70 \\
		MM-2FSK$\Delta$4.0 & 0.53 & 0.35 & 0.41 & 0.37 \\
		MM-2FSK$\Delta$8.0 & 0.54 & 0.22 & 0.24 & 0.21 \\
		MM-2FSK$\Delta$10.0 & \best{0.51} & \best{0.18} & \best{0.19} & \best{0.17} \\
		\bottomrule
	\end{tabularx} 
	
	\label{table:ablation_mm2fsk_frequencies}
\end{table}

We assess the MM-2FSK algorithm \VWedit{in the static scenario} using the frequency configurations outlined in \autoref{table:2fsk_frequencies} to simulate different radar systems.
To isolate the impact of the frequency configuration from sensor characteristics, we utilize per-point depth priors obtained from the ground-truth \VWedit{MVS} measurement system.
The results are summarized in \autoref{table:ablation_mm2fsk_frequencies}, showcasing performance across all four metrics, averaged over the 45 objects of the \VWedit{MAROON} dataset.

\begin{figure}[!htbp]
	\centering
	\includegraphics[width=\linewidth]{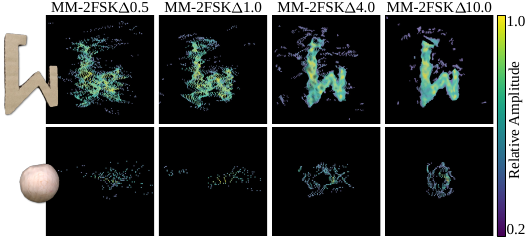} 
	\caption{\VWedit{Static} MM-2FSK reconstructions \VWedit{with depth priors obtained from MVS } for the \obj{Cardboard} and \obj{Wood Ball} objects, compared across different frequency configurations. \VWedit{The displayed point clouds are colored by the relative amplitude $V/V_{\text{max}}$, where $V_{\text{max}}$ is the maximum amplitude per reconstruction. We visualize a high dynamic range of $[-14~\text{dB}, 0~\text{dB}]$, mapped to linear values in $[0.2, 1.0]$. We conclude that} higher band\-widths exhibit fewer artifacts as they are less sensible to noisy phase variations.}
	\label{fig:validation_mm_fsk_frequencies}
\end{figure}

We observe a trend towards better performance at higher frequency differences, with the MM-2FSK$\Delta$10.0 method achieving the best results, yielding reconstruction errors in millimeter range, with a maximum pixel-wise depth error of only $1.9$~mm with respect to {\mthree}.
We suggest this trend is related to the maximum unambiguous depth correction (cf.~\autoref{table:2fsk_frequencies}), which decreases as frequency difference increases, thereby constraining the radar depth variance.
Specifically, the maximum unambiguous depth correction is inversely proportional to the phase sensitivity~\cite{braeunig_2024}, which means that larger frequency differences are less sensitive to phase variations due to clutter and noise.
This phenomenon becomes more evident when visualizing the corresponding point clouds of the reconstructions, as shown in \autoref{fig:validation_mm_fsk_frequencies}.
Reconstructions with higher frequency differences exhibit fewer noise artifacts.

\subsection{\VWedit{Static} Comparison \VWedit{against} the State of the Art}

\begin{figure*}[!htbp]
	\centering
	\includegraphics[width=0.90\linewidth]{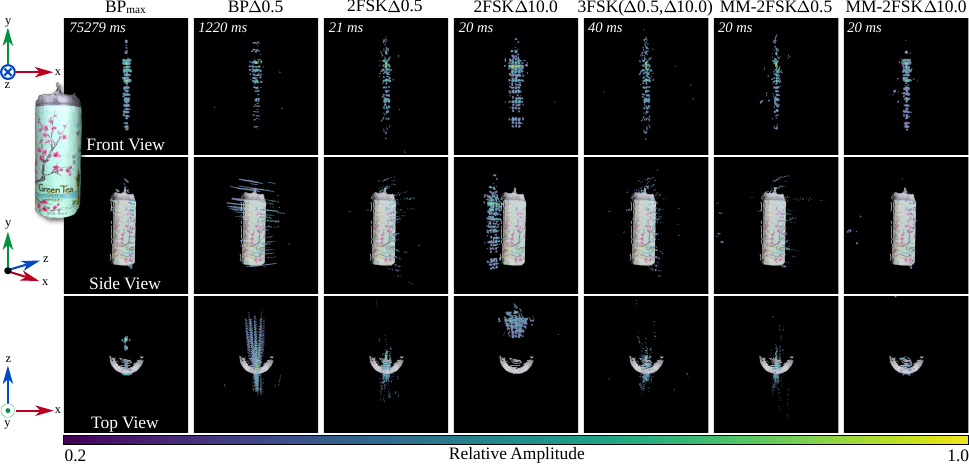} 
	\caption{\VWedit{Static object} comparison of the reconstructed point clouds for backprojection, 2FSK, 3FSK, and MM-2FSK \VWedit{(priors from RealSense)} across different frequency configurations and views, with \VWedit{$\text{BP}_{\text{max}}$} using the full 10~GHz frequency spectrum at 128 frequency steps. 
		\VWedit{The displayed point clouds are colored by the relative amplitude $V/V_{\text{max}}$, where $V_{\text{max}}$ is the maximum amplitude per reconstruction. We visualize a high dynamic range of $[-14~\text{dB}, 0~\text{dB}]$, mapped to linear values in $[0.2, 1.0]$. }
		\VWedit{Furthermore, we denote the reconstruction times of every method in the top left corner of the first row.} 
		Side and top views overlay the reconstructed point cloud with the ground-truth point cloud for the \obj{Bottle} object at 30 cm object-to-sensor distance. The MM-2FSK method exhibits fewer artifacts and is closer to the ground truth than backprojection and 2FSK at the same frequency configuration. Additionally, depending on the frequency difference, MM-2FSK performs as well as or better than 3FSK, which employs a greater number of frequencies.}
	\label{fig:validation_mm_fsk}
\end{figure*}

We compare the performance of MM-2FSK against 2FSK, 3FSK and BP, using per-point depth priors obtained from the \VWedit{RealSense} active stereo depth camera.
Our evaluation focuses on three frequency configurations from \autoref{table:2fsk_frequencies}: the first, where $\widetilde{d}$ lies within the maximum unambiguous depth correction ($\Delta f = 0.5$), the second, where it narrowly exceeds this factor ($\Delta f = 1.0$), and finally, the configuration where MM-2FSK performed best in previous ablation ($\Delta f = 10.0$).
Additionally, we present reference radar reconstructions derived from the significantly more resource-intensive backprojection $\text{BP}_{\text{max}}$, utilizing the maximum of 128 frequency steps.

\begin{table}[!bp]
	\caption{\VWedit{Quantitative evaluation against the state of the art.} The mean reconstruction error is given in centimeters, evaluated for backprojection, 2FSK and, MM-2FSK \VWedit{(priors from RealSense)} against the ground-truth setup. Each metric is averaged across all \VWedit{static} MAROON objects at a sensor distance of 30~cm. The best results per metric among all high-speed imaging methods are highlighted.}
	\begin{tabularx}{\linewidth}{@{}cYYYY@{}} 
		\toprule 
		& { \mone } & { \mtwo }  & { \mthree }  & { \mfour } \\ 
		\midrule
		$\text{BP}_\text{max}$     & 0.82 & 0.90  & 0.94  & 0.85 \\
		\midrule
		BP$\Delta$0.5 & \best{0.69} &  4.27  & 3.34  & 3.18  \\
		BP$\Delta$1.0 & 0.79 & 5.29 & 3.70 & 3.40 \\
		BP$\Delta$10.0 & 1.02 & 6.35 & 4.98 & 4.87 \\
		\midrule
		2FSK$\Delta$0.5 & 0.90 & 2.36 & 2.55  & 2.10 \\
		2FSK$\Delta$1.0 & 8.68 & 12.27 & 12.87 & 12.82 \\
		2FSK$\Delta$10.0 & 9.80 & 9.69 & 10.38 & 10.38 \\
		\midrule
		3FSK($\Delta$0.5,$\Delta$10.0) & 0.83 & 2.54 & 2.63  & 2.30 \\
		3FSK($\Delta$1.0,$\Delta$10.0) & 8.37 & 12.47 & 12.88 & 12.87 \\
		3FSK($\Delta$2.0,$\Delta$10.0) & 6.43 & 7.90 & 8.51 & 8.52 \\
		\midrule
		\midrule
		MM-2FSK$\Delta$0.5 & 0.95 & 2.95  & 3.13 & 2.39 \\
		MM-2FSK$\Delta$1.0 & 0.98 & 2.72 & 2.84 & 2.28 \\
		MM-2FSK$\Delta$10.0 & 0.82 & \best{1.74} & \best{1.36} & \best{1.01} \\
		\bottomrule
	\end{tabularx}
	\label{table:validation_mm_fsk}
\end{table}

In \autoref{fig:validation_mm_fsk}, we present exemplary reconstructions of the \obj{Bottle} object generated by backprojection, 2FSK, 3FSK, and MM-2FSK.
Among these methods with the same frequency configuration, MM-2FSK is the closest to the ground truth; at $\Delta f=10.0$~GHz (\textit{rightmost column}), it avoids reconstructing the partially transmissive plastic, resulting in a closer match to the ground-truth reconstruction than $\text{BP}_{\text{max}}$.

The qualitative observations align with the quantitative evaluations in \autoref{table:validation_mm_fsk}, which detail the reconstruction errors in centimeters across all objects; here, MM-2FSK$\Delta$10.0 outperforms all approaches of similar number of frequencies across three metrics.
Additionally, its performance in pixel-wise depth estimation, measured by metrics {\mthree} and {\mfour}, approaches that of $\text{BP}_{\text{max}}$, with only +4.2~mm and +1.6~mm additional error relative to the ground truth, respectively, despite utilizing significantly fewer frequencies.

In terms of different frequency configurations, our method proves to be more robust than other approaches, either matching or exceeding their performance.
In contrast, both 2FSK and 3FSK show significantly poorer results when the depth prior falls outside the maximum unambiguous depth correction range, particularly when $\Delta f > 0.5$~GHz.

\begin{figure*}[!htbp]
	\centering
	\includegraphics[width=\linewidth]{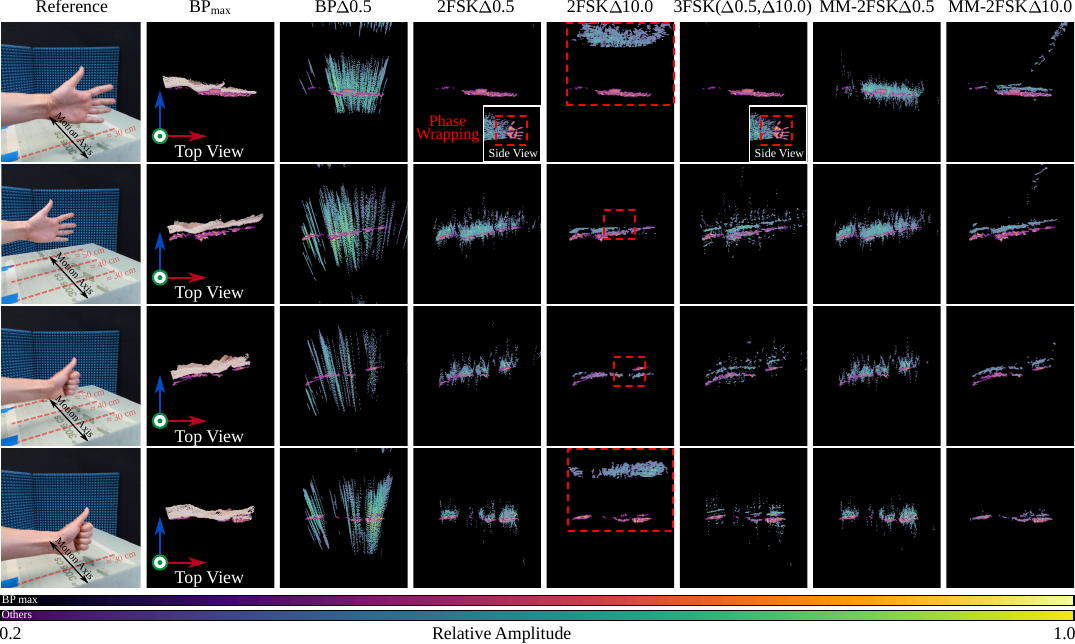} 
	\caption{\VWedit{Qualitative results of two dynamic sequences with radial hand motion, where reference views originate from Azure Kinect RGB frames, with dashed red overlays marking approximate target-to-radar ranges; the 40 cm range aligns most closely with the scalar depth priors used in 2FSK and 3FSK. We present top-down views of the reconstructed point clouds using backprojection, 2FSK, 3FSK, and MM-2FSK (priors from Azure Kinect) across different frequency configurations. 
		We employ a distinct colormap for $BP_{\text{max}}$ to differentiate its measurements from few‑tone reconstructions. We overlay $BP_{\text{max}}$ with per‑point Azure Kinect measurements (\textit{Second column from left}), highlighting data mismatch. Furthermore, we overlay few-tone re\-con\-struc\-tions with $BP_{\text{max}}$ to assess relative depth deviation. 
		Across all few-tone settings, MM-2FSK shows the closest agreement with $BP_{\text{max}}$, exhibiting reduced noise artifacts and preserving the reconstructed hand geometry without (partial) phase-wrapping-induced distortions (\textit{red dashed boxes}).
		Within MM-2FSK variants (\textit{two rightmost columns}), a trade-off is observed between reconstructions with lower noise levels (MM-2FSK$\Delta$10.0) and increased robustness to inter-modality depth mismatch due to higher maximum unambiguous depth range (MM-2FSK$\Delta$0.5).
	}}
	\label{fig:dynamic_validation_mm_fsk}
\end{figure*}

\subsection{\VWedit{Dynamic Comparison against the State of the Art}}
\VWedit{
We evaluate two radial hand-motion captures, comparing 2FSK, 3FSK, and BP against MM-2FSK, using time-synchronized depth priors from the Azure Kinect camera. 
All metrics are reported relative to $\text{BP}_{\text{max}}$, established as our best proxy for ground truth (see~\autoref{table:validation_mm_fsk}). 

In~\autoref{table:dynamic_validation_mm_fsk}, we list the mean reconstruction error averaged over five frames of each hand-motion sequence, sampled at timestamps corresponding to 0~\%, 25~\%, 50~\%, 75~\%, and 95~\% of the overall sequence time. 
In the absence of a labeled foreground mask for the human hand, we omit computation of the \mfour~metric. 
Delivering the best results across all metrics, MM-2FSK reconstruction quality is the closest to $\text{BP}_{\text{max}}$ among all few-tone imaging methods.
On the other hand, 2FSK and 3FSK become increasingly difficult to adapt to dynamic scenarios, leading to larger depth deviations from $\text{BP}_{\text{max}}$ than in static experiments.}

\begin{table}[!bp]
	\caption{\VWedit{Quantitative evaluation against the state of the art in a dynamic setting. The mean reconstruction error is given in centimeters, evaluated for backprojection, 2FSK and, MM-2FSK (priors from Azure Kinect) against $\text{BP}_{\text{max}}$. Each metric is averaged across five frames of each radial hand motion sequence. The best results per metric among all high-speed imaging methods are highlighted.}}
	\begin{tabularx}{\linewidth}{@{}cYYYY@{}} 
		\toprule 
		& { \mone } & { \mtwo }  & { \mthree } \\ 
		\midrule
		BP$\Delta$0.5 & 0.37 &  15.12  & 3.89  \\
		BP$\Delta$1.0 & 0.51 & 15.71 & 5.67 \\
		BP$\Delta$10.0 & 0.34 & 15.16 & 5.23 \\
		\midrule
		2FSK$\Delta$0.5 & 2.48 & 3.80 & 8.01 \\
		2FSK$\Delta$1.0 & 3.97 & 4.92 & 9.52 \\
		2FSK$\Delta$10.0 & 5.43 & 5.61 & 11.61 \\
		\midrule
		3FSK($\Delta$0.5,$\Delta$10.0) & 2.24 & 3.98 & 3.75 \\
		3FSK($\Delta$1.0,$\Delta$10.0) & 3.88 & 5.21 & 8.46 \\
		3FSK($\Delta$2.0,$\Delta$10.0) & 4.43 & 5.60 & 10.21 \\
		\midrule
		\midrule
		MM-2FSK$\Delta$0.5 & 0.28 & 2.36 & \textbf{1.01} \\
		MM-2FSK$\Delta$1.0 & \textbf{0.21} & 1.77 & 4.47 \\
		MM-2FSK$\Delta$10.0 & 0.59 & \textbf{1.59} & 3.84 \\
		\bottomrule
	\end{tabularx}
	\label{table:dynamic_validation_mm_fsk}
\end{table}

\VWedit{In~\autoref{fig:dynamic_validation_mm_fsk}, we present additional reconstructions of the two radial hand motions.
Consistent with the experiments in a static setting, MM‑2FSK is more robust than other few‑tone methods, exhibiting lower noise and fewer phase‑wrapping errors.
We also observe the expected frequency configuration trade‑offs from prior ablations: a smaller frequency difference $\Delta f$ increases noise but yields a higher maximum unambiguous depth correction, enabling greater tolerance to uncertainty in the depth prior.
This is evident in the two rightmost columns of \autoref{fig:dynamic_validation_mm_fsk}: MM-2FSK$\Delta10.0$ shows misalignment due to data mismatch between the primary and secondary modality, whereas MM-2FSK$\Delta0.5$ includes more points overlapping with $\text{BP}_{\text{max}}$  at the cost of higher noise.
These results suggest that the frequency configuration should be selected not only according to spatial but also temporal mismatch when fusing modalities into a common reference frame.
}

\section{Discussion}
Our experiments demonstrate that the proposed MM-2FSK algorithm outperforms comparable radar-only algorithms.
While integrating a \VWedit{secondary modality} yields superior results, it also renders the algorithm prone to \VWedit{sensing limitations of this modality; in the case of an optical sensor, representative limitations may arise from unsuitable lighting conditions or highly reflective materials.}
Although our triangulation method may still provide reasonable depth values to fill in missing depth priors, an alternative approach could involve fusing the 3FSK method with our work, provided that the radar sensor supports high bandwidth and signal modulation at non-equidistant frequency steps.

Furthermore, the triangulation method does not respect object boundaries, such that in complex scenarios with multiple surface targets, depth interpolation using triangle topology may yield insufficient depth priors.
An interesting future task could be the incorporation of semantic knowledge about the environment, as achieved by object segmentation. %

\VWedit{Based on our experiments with optical depth sensors, } we \VWedit{further} recognize that sensor fusion with \VWedit{complementary modalities} limits radar-specific characteristics, such as signal transmission, which is desirable in applications like security scanning or medical imaging. 
An intriguing \VWedit{extension of our work} would be to investigate \VWedit{assistive modalities} with transmission properties \VWedit{closer to the primary modality, for example from the Terahertz frequency bands}.
Ultimately, it is essential to carefully evaluate the trade-off between the desirable characteristics of the radar sensor and the constraints imposed by the supporting \VWedit{modality} for each application individually.

\section{Conclusion}
In this work, we address the increasing demand for radar sensors capable of high-speed capture and reconstruction to enable fast and accurate depth sensing of both static and dynamic targets by presenting a novel multimodal signal processing method based on frequency shift keying principles for \VWedit{near-field mmWave} MIMO radar imaging~\cite{braeunig_2023_fsk}. 

Leveraging the capabilities of an assistive \VWedit{secondary modality}, our proposed MM-2FSK algorithm overcomes current limitations of the 2FSK~\cite{braeunig_2023_fsk} approach with respect to the maximum unambiguous depth correction factor. 
\VWedit{By incorporating geometry processing into the reconstruction pipeline, we exploit additionally captured depth maps to create per-point depth priors from the perspective of a near-field, high-resolution radar. 
Collectively, these geometry-inspired approaches reduce the risk of erroneous measurements and range-phase ambiguities during MM-2FSK processing.}
This simple yet effective approach allows us to generalize our MM-2FSK extension to capture environments where neither the object's position nor its geometry is known in advance.

\VWedit{We evaluate the proposed method on a diverse set of static objects in the MAROON dataset and on dynamic radial hand-motion captures, using a high-resolution MIMO radar in conjunction with multiple secondary sensors from the optical frequency spectrum: multi-view stereo, active stereo, and optical time-of-flight. }
Our results demonstrate that our multimodal imaging approach outperforms comparable related work in terms of depth quality and performs only marginally worse than backprojection with maximum frequency steps, despite using fewer frequencies, therefore significantly lowering the capture and computation time.
In summary, we believe our method holds great potential for future applications in multi-sensor target tracking.

\section*{Acknowledgement}
The authors would like to express their gratitude to Paul Himmler for the insightful discussions.

This work was funded by the Deutsche Forschungsgemeinschaft (DFG, German Research Foundation) – SFB 1483 – Project-ID 442419336, EmpkinS.

The authors would like to thank the Rohde \& Schwarz
GmbH \& Co. KG (Munich, Germany) for providing the radar
imaging devices. 

The authors gratefully acknowledge the scientific support and HPC resources provided by the Erlangen National High Performance Computing Center of the Friedrich-Alexander-Universität Erlangen-Nürnberg.

\bibliographystyle{ieeetr}
\bibliography{Bibliography}
\end{document}